\documentclass[conference, 10pt]{IEEEtran} 

\usepackage[utf8]{inputenc} 
\usepackage[T1]{fontenc}    
\usepackage{booktabs}       
\usepackage{graphicx}
\usepackage{nicefrac}       
\usepackage{microtype}      
\usepackage[cmex10]{amsmath}
\usepackage{amssymb}
\usepackage{amsfonts}
\usepackage{thmtools}
\usepackage{cite}
\usepackage{subcaption}
\usepackage{tikz}
\usepackage[linesnumbered,ruled]{algorithm2e}

\makeatletter
\newcommand{\removelatexerror}{\let\@latex@error\@gobble}
\makeatother

\usepackage{color}

\definecolor{Rafael}{RGB}{139,0,0} 

\definecolor{rick}{RGB}{0,139,0}

\definecolor{old}{RGB}{0,0,139}

\title{Deep Learning for Channel Coding via Neural Mutual Information Estimation}

\author{
	  \authorblockN{Rick Fritschek$^{\ast}$, Rafael F. Schaefer$^{\dagger}$, and Gerhard Wunder$^{\ast}$\\[2mm]}
	  \IEEEauthorblockA{\begin{tabular}{cc}
	       \begin{tabular}{c}
	           $^{\ast}$ Heisenberg Communications and Information Theory Group\\
                        Freie Universit\"at Berlin, \\
                        Takustr. 9,
                        14195 Berlin, Germany\\
                        Email: \texttt{\{rick.fritschek, g.wunder\}@fu-berlin.de}
	       \end{tabular}
	       \begin{tabular}{c}
	           $^{\dagger}$ Information Theory and Applications Chair\\
	                        Technische Universit{\"a}t Berlin \\
                            Einsteinufer 25, 10587 Berlin, Germany\\
                            Email: \texttt{rafael.schaefer@tu-berlin.de}
	       \end{tabular}
	  \end{tabular}
}}

\begin{document}

\maketitle

\begin{abstract}
End-to-end deep learning for communication systems, i.e., systems whose encoder and decoder are learned, has attracted significant interest recently, due to its performance which comes close to well-developed classical encoder-decoder designs. However, one of the drawbacks of current learning approaches is that a differentiable channel model is needed for the training of the underlying neural networks. In real-world scenarios, such a channel model is hardly available and often the channel density is not even known at all. Some works, therefore, focus on a generative approach, i.e., generating the channel from samples, or rely on reinforcement learning to circumvent this problem. We present a novel approach which utilizes a recently proposed neural estimator of mutual information. We use this estimator to optimize the encoder for a maximized mutual information, only relying on channel samples. Moreover, we show that our approach achieves the same performance as state-of-the-art end-to-end learning with perfect channel model knowledge.
\end{abstract}

\section{Introduction}

Deep learning based methods for wireless communication is an emerging field whose performance is becoming competitive to state-of-the-art techniques that evolved over decades of research. One of the most prominent recent examples is the end-to-end learning of communication systems utilizing (deep) neural networks (NNs) as encoding and decoding functions with an in-between noise layer that represents the channel \cite{OShea2017}. In this configuration, the system resembles the concept of an autoencoder in the field of machine learning, which does not compress but adds redundancy to increase reliability. These encoder-decoder systems can achieve bit error rates which come close to practical baseline techniques if they are used for over-the-air transmissions \cite{SBrink2018}. This is promising since complex encoding and decoding functions can be learned on-the-fly without extensive communication-theoretic analysis and design, possibly enabling future communication systems to better cope with new and changing channel scenarios and use-cases. However, the previously mentioned approaches have the drawback that they require a known channel model to properly choose the noise layer within the autoencoder. Moreover, this channel model needs to be differentiable to enable back-propagation through the whole system to optimize, i.e., learn, the optimal weights of the NN. 

One approach is to assume a generic channel model, e.g. a Gaussian model. The idea is then to first learn according to this general model and subsequently fine-tune the receiver, i.e., the weights of the decoding part of the NN, based on the actual received signals. This approach was implemented in \cite{SBrink2018}. Another approach is to use generative adversarial networks (GANs), which was introduced in \cite{goodfellow2014generative}. GANs are composed of two competing neural networks, i.e., a generative NN and a discriminative NN. Here, the generative NN tries to transform a uniform input to the real data distribution, whereas the discriminative NN compares the samples of the real distribution (from the data) to the fake generated distribution and tries to estimate the probability that a sample came from the real data. Therefore, both neural networks are competing against each other. Due to their effectiveness and good performance, GANs are a popular and active research direction. In our problem of end-to-end learning for communications, GANs were used in \cite{ye2018GAN,oShea2018GAN} to produce an artificial channel model which approximates the true channel distribution and can therefore be used for end-to-end training of the encoder and decoder. The third approach is using reinforcement learning (RL) and is, therefore, circumventing the problem with the back-propagation itself by using a feedback link \cite{aoudia2018end}. In this approach, the transmitter can be seen as an agent which performs actions in an environment and receives a reward through the feedback link. The transmitter can then be optimized to minimize an arbitrary loss function, connected to the actions, i.e., the transmitted signals in our case. This work was subsequently extended towards noisy feedback links in \cite{goutay2018deep}. However, the drawback of this approach is, that RL is known for its sample inefficiency, meaning that it needs large sample sizes to achieve high accuracy. Moreover, both approaches (via RL and GAN) still have a dependence on the receiver of the system. As the GAN approach approximates the channel density as a surrogate for the missing channel model, it still needs end-to-end learning in the last step. The RL approach, on the other hand, needs the feedback of the receiver and therefore also depends on the decoder. In this paper, we make progress on this by proposing a method that is completely independent of the decoder. This circumvents the challenge of a missing channel model.

\begin{figure}
    \centering
    \includegraphics[scale=0.97]{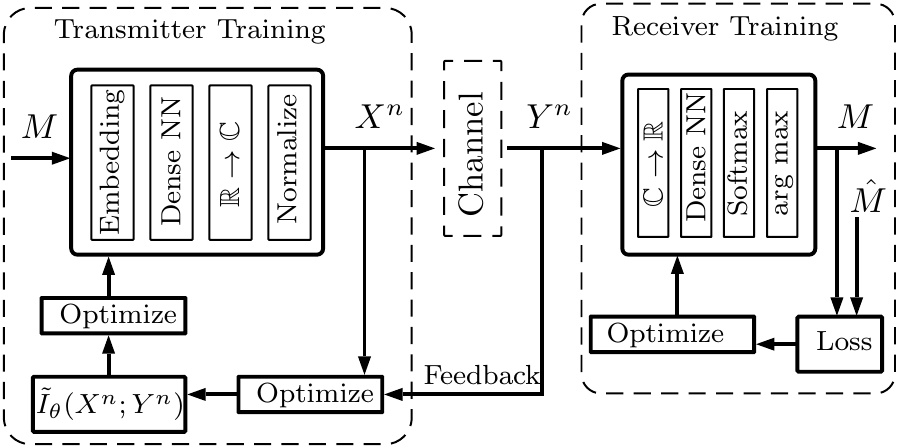}
    \caption{The figure shows our channel coding approach. We train an approximation of the mutual information $\tilde{I}_{\theta}(X^n;Y^n)$ between the channel input and output samples ($x^n,y^n$), which is used to optimize the neural network of the channel encoder. Both are alternatingly trained until convergence.}
    \label{fig:my_label}
\end{figure}

{\bf Our contribution:} 
From a communication theoretic perspective, we know that the optimal transmission rate is a function of the mutual information $I(X;Y)$ between input $X$ and output $Y$ of a channel $p(y|x)$. For example, the capacity $C$ of an additive white Gaussian noise (AWGN) channel (as properly defined in the next section) is given by the maximum of the mutual information over the input distribution $p(x)$ under an average power constraint $P$, i.e., 
\begin{equation}
    C = \max_{p(x): \mathbb{E}(X^2)\leq P} I(X;Y) = \log\Big(1+\frac{P}{\sigma^2}\Big).
\end{equation}
This suggests to use mutual information as a metric to learn the optimal channel encoding function of AWGN channels as well as other communication channels. However, the mutual information also dependents on the channel probability distribution. But instead of approximating the channel probability distribution itself, we will approximate the mutual information between the samples of the channel input and output and optimize the encoder weights, by maximizing the mutual information between them, see Fig.~\ref{fig:my_label}. For that, we utilize a recent NN estimator of the mutual information \cite{belghazi2018mine} and integrate it in our communication framework. We are, therefore, independent of the decoder and can reliably train our encoding function using only channel samples.

{\bf Notation:} We stick to the convention of upper case random variables $X$ and lower case realizations $x$, i.e. $X\sim p(x)$, where $p(x)$ is the probability mass or density function of $X$. Moreover, $p(x^n)$ is the probability mass or density function of the random vector $X^n$. We also use $|\mathcal{X}|$ to denote the cardinality of a set $\mathcal{X}$. The expectation is denoted by $\mathbb{E}[\cdot]$.

\section{Point-to-point Communication Model}
\label{Model_Section}
We consider a communication model with a transmitter, a channel, and a receiver. The transmitter wants to send a message $m\!\in\! \mathcal{M}\!=\!\{1,2, \ldots, 2^{nR}\}$ at a rate $R$ over a noisy channel using an encoding function $f(m)=x^n(m)\in\mathbb{C}^n$ to make the transmission robust against noise. Moreover, for every message $m\in\mathcal{M}$ we assume an average power constraint $\tfrac{1}{n}\sum_{i=1}^n|x_i(m)|^2\leq P$ on the corresponding codewords. In this work, our communication channel is an AWGN channel such that the received signal is given as 
\begin{equation}
    Y_i=X_i+Z_i,\qquad \mbox{for } i\in \{1,\ldots, n\}
\end{equation}where the noise $Z_i$ is i.i.d. over $i$ with $Z_i\sim \mathcal{CN}(0,\sigma^2)$. The receiver uses a decoder $g(y^n)=\hat{m}$ to estimate and recover the original message. Moreover, the block error rate $P_e$ is defined as the average probability of error over all messages
\begin{equation}
    P_e= \frac{1}{|\mathcal{M}|}\sum_{m=1}^{|\mathcal{M}|} \mbox{Pr}(\hat{M}\neq m | M = m).
\end{equation}
\section{Neural Estimation of Mutual Information}

A straight forward computation and therefore evaluation of the mutual information is difficult due to its dependence on the joint probability density of the underlying random variables. A fallback solution is therefore a limitation on mutual information approximations. The main challenge here is to provide an accurate and stable approximation from low sample sizes. Common approaches are based for example on binning of the probability space \cite{fraser1986independent,darbellay1999estimation}, $k$-nearest neighbor statistics \cite{kraskov2004estimating,gao2015efficient,GaoEstimatorMI}, maximum likelihood estimation \cite{suzuki2008approximating}, and variational lower bounds \cite{barber2003algorithm}. We focus on a recently proposed estimator \cite{belghazi2018mine}, coined \emph{mutual information neural estimation (MINE)}, which utilizes the Donsker-Varadhan representation of the Kullback-Leibler divergence, which in turn is connected to the mutual information by \begin{IEEEeqnarray}{rCl}
    I(X;Y)&:=&\int_{\mathcal{X} \times \mathcal{Y}} p(x,y) \log \frac{p(x,y)}{p(x) p(y)} dxdy\IEEEnonumber\\
    &=&D_{KL}(p(x,y)||p(x) p(y))\IEEEnonumber\\\IEEEnonumber
    &=&\mathbb{E}_{p(x,y)}\left[\log \frac{p(x,y)}{p(x) p(y)} \right].
\end{IEEEeqnarray}

The Donsker-Varadhan representation can be stated as 
\begin{equation}
    D_{KL}(P||Q)=\sup_{g:\Omega \rightarrow \mathbb{R}} \mathbb{E}_P[g(X,Y)]-\log( \mathbb{E}_Q[e^{g(X,Y)}]) \label{DV-rep}
\end{equation}
where the supremum is taken over all measurable functions $g$ such that the expectation is finite. Now, depending on the function class, the right hand side of \eqref{DV-rep} yields a lower bound on the KL-divergence, which is tight for optimal functions. In \cite{belghazi2018mine} Belghazi \emph{et al.} proposed to choose a neural network, parametrized with $\theta \in \Theta$ as function family $T_\theta: \mathcal{X} \times \mathcal{Y}\rightarrow \mathbb{R}$ for the lower bound. This yields the estimator 
\begin{equation}
    I(X;Y)\geq \sup_{\theta \in \Theta} \mathbb{E}_{p(x,y)}^{}[T_\theta(X,Y)]-\log \mathbb{E}^{}_{p(x)p(y)}[e^{T_\theta(X,Y)}].\label{DVapprox}
\end{equation}
Moreover, they show that the above estimator is \emph{consistent} in the sense that it converges to the true value for increasing sample size $k$. Another closely related estimator is based on $f$-divergence representations \cite{nguyen2010estimating} and was recently applied for $f$-GANs \cite{nowozin2016f}, which uses the Fenchel duality to bound the $f$-divergence from below as 
\begin{equation}
    D_f(P||Q) \geq \sup_{g: \Omega \rightarrow \mathbb{R}} \mathbb{E}_P[g(X,Y)]-\mathbb{E}_Q[f^*(g(X,Y))]
\end{equation} 
where the supremum is over all measurable functions $g$. Moreover, \cite{nowozin2016f} also proposed to choose a parameterized neural network for this function family and provide a table for the right choice of the conjugate dual function $f^*$, which is $\exp(x-1)$ to obtain a lower bound on the KL-divergence. This leads to the estimator 
\begin{equation}
    I(X;Y) \geq \sup_{\theta \in \Theta} \mathbb{E}_{p(x,y)}[T_\theta(X,Y)]-\mathbb{E}_{p(x)p(y)}[e^{ T_\theta(X,Y)-1}].\label{NGYapprox}
\end{equation}
We note that both estimators can be derived through application of the Fenchel duality. Moreover, both lower bounds share the same supremum, however, over the choice of functions $T$, \eqref{DVapprox} is closer to the supremum than \eqref{NGYapprox}, see \cite{ruderman2012tighter}. The work of \cite{belghazi2018mine} compares both approximations and shows that \eqref{DVapprox} provides a tighter estimate for high-dimensional variables. We therefore focus on the latter in our following implementation.

\section{Implementation}
\subsection{Encoder Training via Mutual Information Estimation}
\label{Training_Encoder}
\def\layersep{2cm}
\tikzset{%
  neuron missing/.style={
    draw=none, 
    scale=4,
    text height=0.333cm,
    execute at begin node=\color{black}$\vdots$
  },
}

\begin{figure}
    \centering
    \begin{tikzpicture}[shorten >=1pt,->,draw=black!50, node distance=\layersep,scale=0.8, every node/.style={scale=0.8}]
    \tikzstyle{every pin edge}=[<-,shorten <=1pt]
    \tikzstyle{neuron}=[circle,fill=black!25,minimum size=17pt,inner sep=0pt]
    \tikzstyle{dots}=[draw=none,scale=1.5,text height=0.2cm,execute at begin node=\color{black}$\vdots$];
    \tikzstyle{input neuron}=[neuron, fill=green!50];
    \tikzstyle{output neuron}=[neuron, fill=red!50];
    \tikzstyle{hidden neuron}=[neuron, fill=blue!50];
    \tikzstyle{annot} = [text width=5em, text centered]

    
        \node[input neuron] (I-1) at (0,-1.5) {$x_1$};
        \node[input neuron] (I-3) at (0,-3) {$x_n$};
    \foreach \name / \x in {2}
        \node[dots, pin=left:Input $X^n$] (Dx-\name) at (0,-2.3) {};
    
        \node[input neuron] (I-4) at (0,-4) {$y_1$};
         \node[input neuron] (I-6) at (0,-5.5) {$y_n$};
    \foreach \name / \y in {5}
        \node[dots,pin=left:Input $Y^n$] (Dy-\name) at (0,-4.8) {};
        
    \foreach \name / \x in {1,2,3,4,6}
            \node[hidden neuron] (H1-\name) at (\layersep,-\x cm) {};
    \foreach \name / \x in {5}
        \node[dots] (Dh1-\name) at (\layersep,-\x) {};
    \foreach \name / \x in {1,2,3,4,6}
            \node[hidden neuron] (H2-\name) at (2*\layersep,-\x cm) {};
    \foreach \name / \x in {5}
        \node[dots] (Dh2-\name) at (2*\layersep,-\x) {};
    
    \node[output neuron, pin={[pin edge={->}]right:$T_{\theta}(X^n,Y^n)$}, right of=H2-3] (O) at (2*\layersep,-3.5) {};

    \foreach \source in {1,3,4,6}
        \foreach \dest in {1,2,3,4,6}
            \path (I-\source) edge (H1-\dest);
    \foreach \source in {1,2,3,4,6}
        \foreach \dest in {1,2,3,4,6}
            \path (H1-\source) edge (H2-\dest);

    \foreach \source in {1,2,3,4,6}
        \path (H2-\source) edge (O);

    \node[annot,above of=H1-1, node distance=1cm] (hl) {Hidden layer (ReLu)};
    \node[annot,above of=H2-1, node distance=1cm] (hl2) {Hidden layer (ReLu)};
    \node[annot,left of=hl] {Input};
    \node[annot,right of=hl2] {Output (linear)};
\end{tikzpicture}
    \caption{Neural network representation of our approximation function $T_\theta$. The samples of $X^n$ and $Y^n$ are concatenated and fed into the network. The network is comprised of two hidden layers, with 20 nodes and ReLu activation function. The output function is linear.}
    \label{fig:MI_neuralEstimation}
\end{figure}
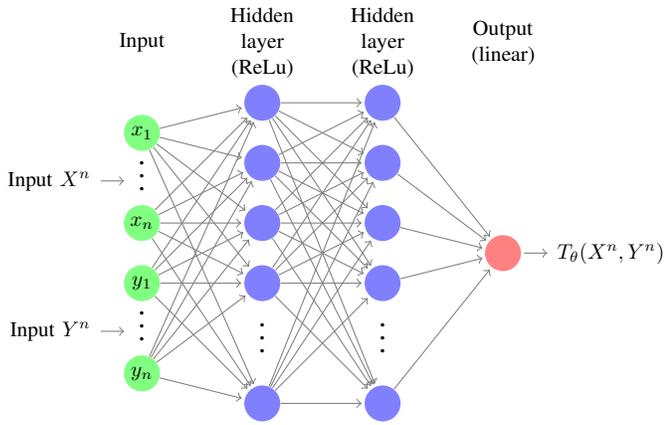

\begin{figure}
    \centering
\begin{subfigure}{.124\textwidth}
  \centering
  \includegraphics[width=\linewidth]{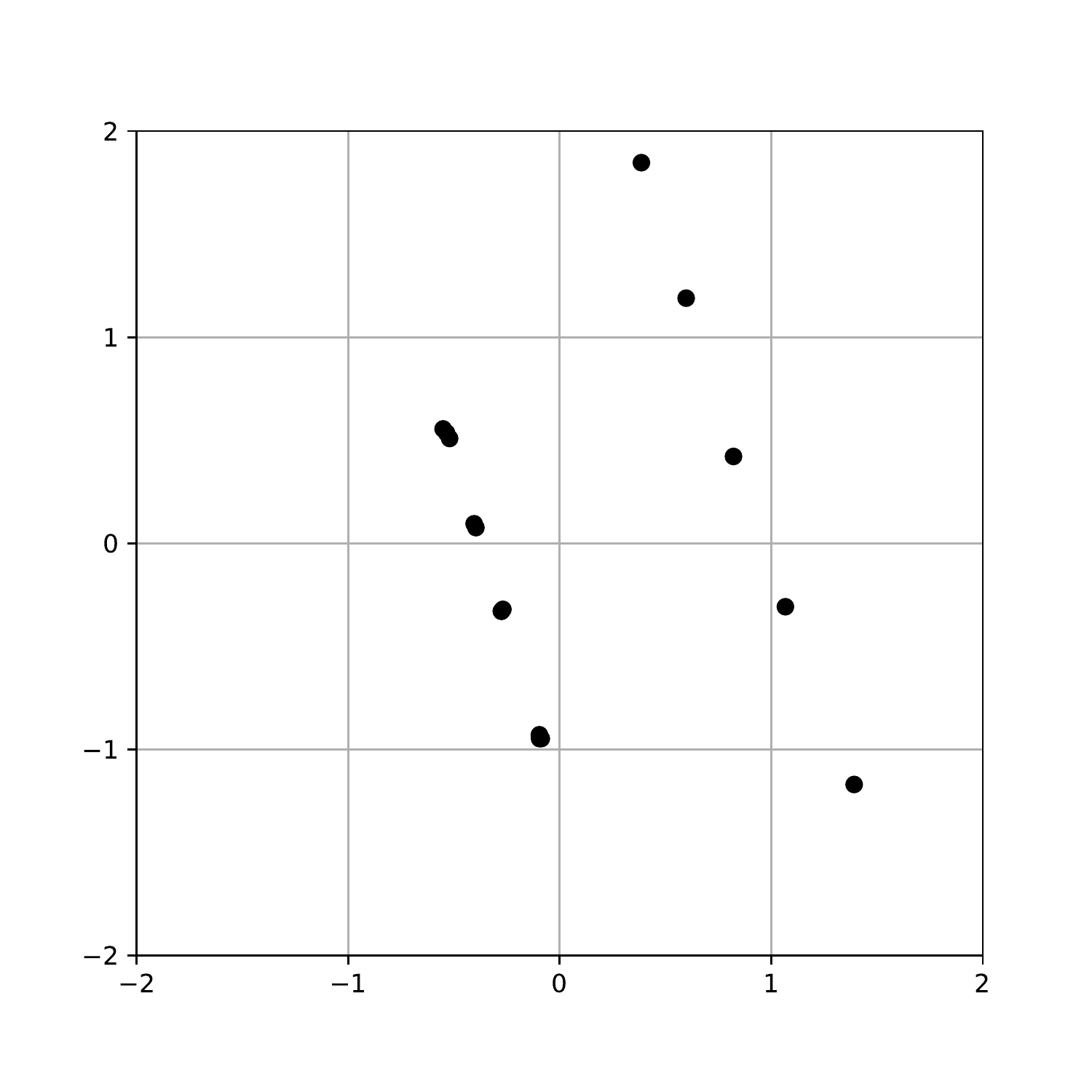}
  \caption{}
  \label{fig:sub1}
\end{subfigure}%
\begin{subfigure}{.125\textwidth}
  \centering
  \includegraphics[width=\linewidth]{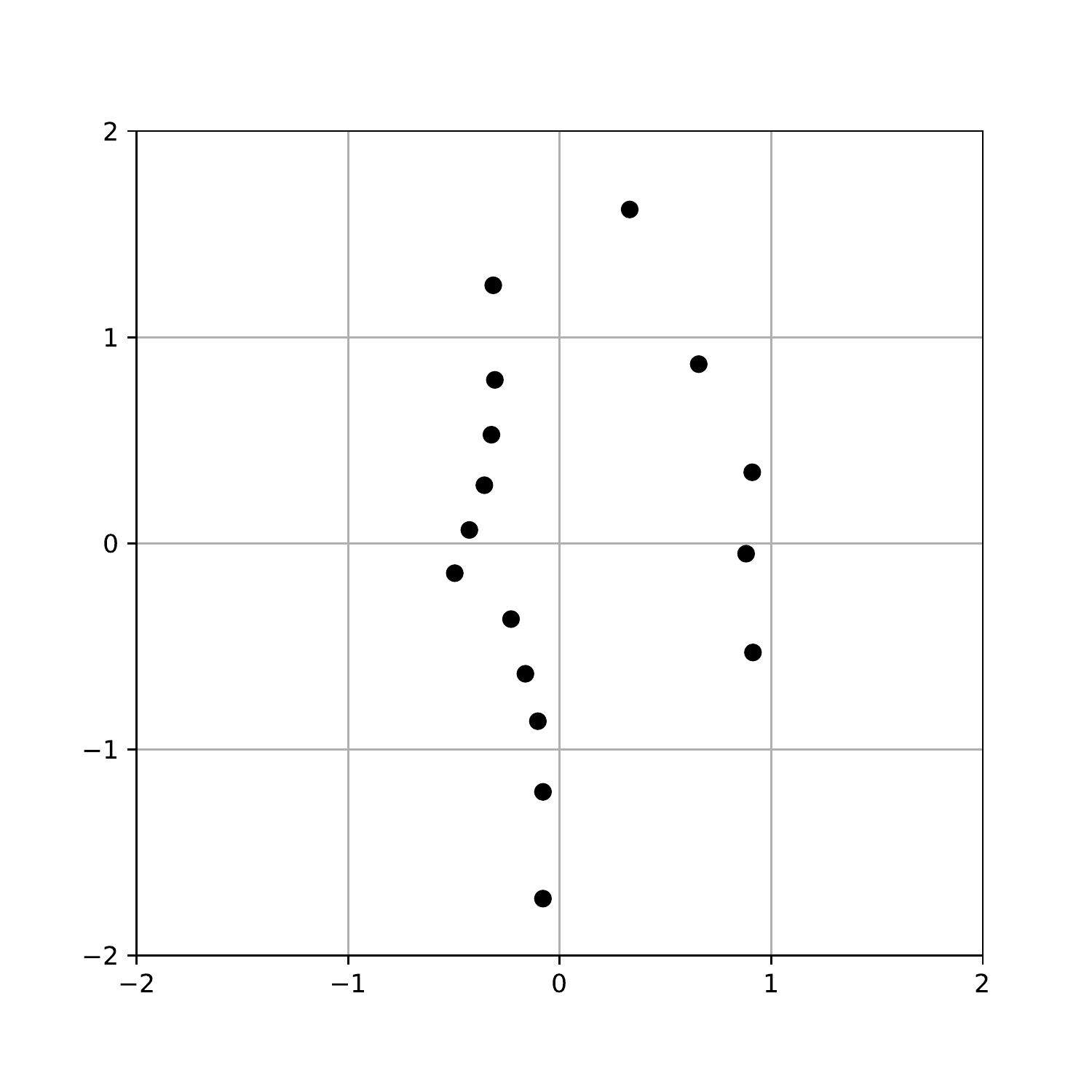}
  \caption{}
  \label{fig:sub2}
\end{subfigure}%
\begin{subfigure}{.125\textwidth}
  \centering
  \includegraphics[width=\linewidth]{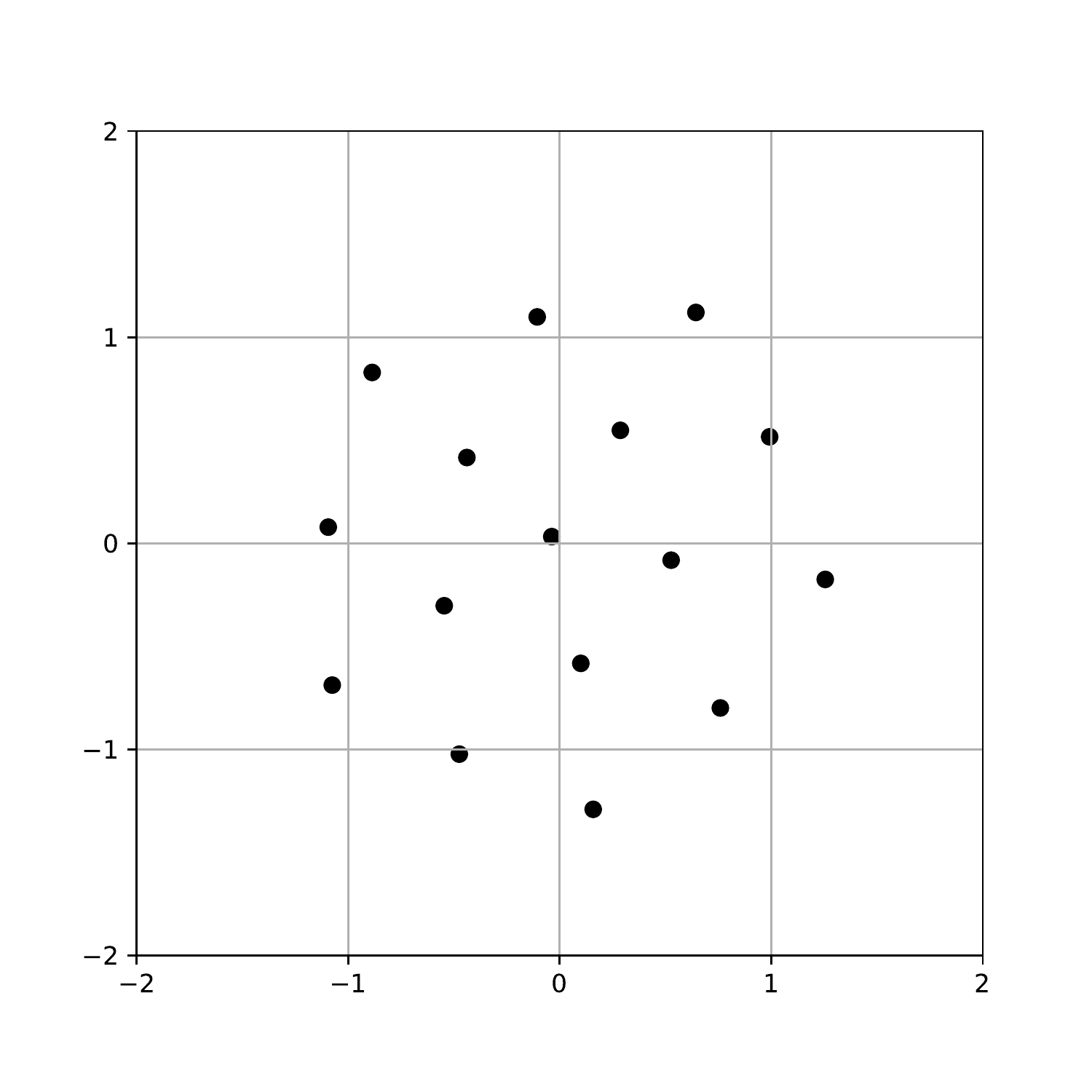}
  \caption{}
  \label{fig:sub3}
\end{subfigure}%
\begin{subfigure}{.125\textwidth}
  \centering
  \includegraphics[width=\linewidth]{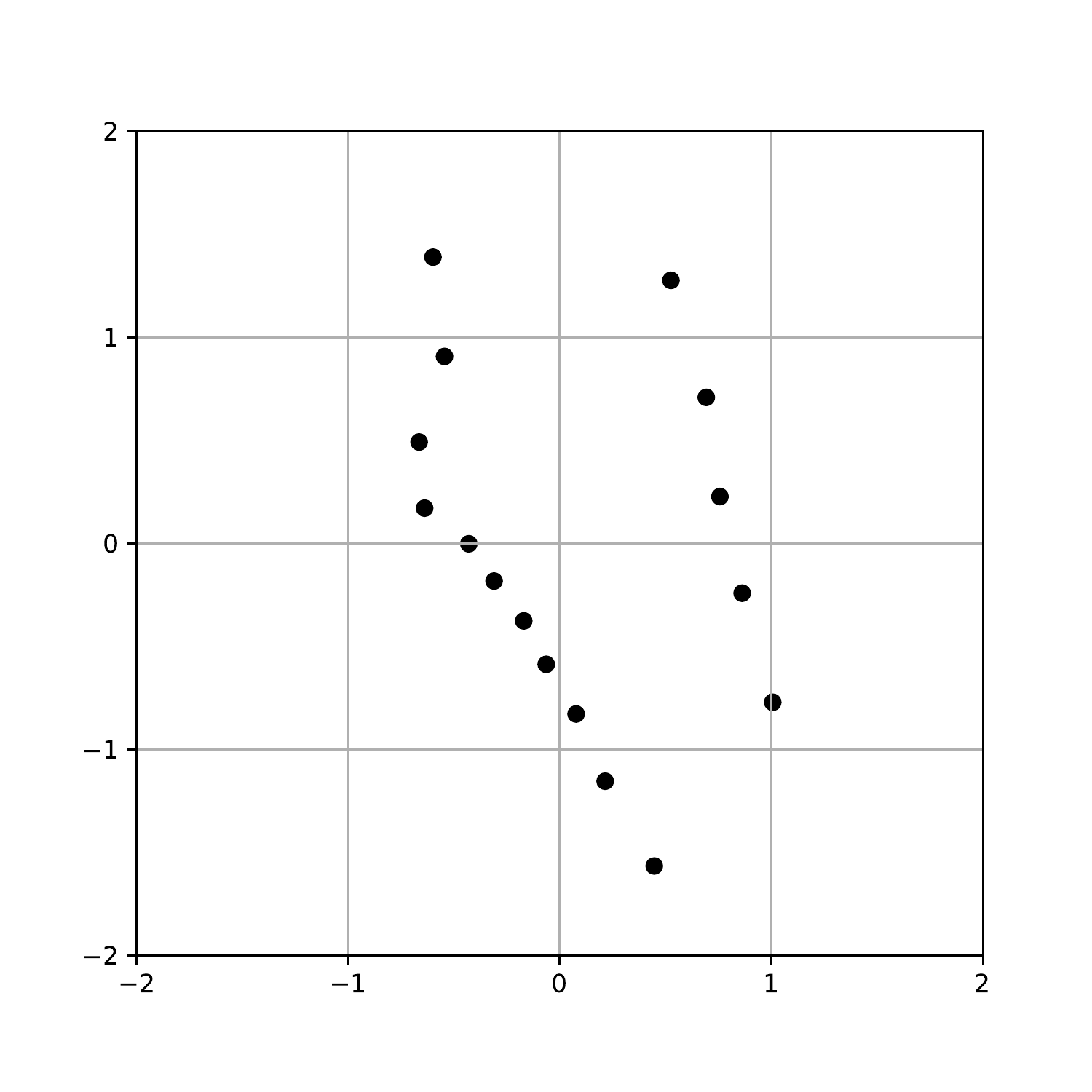}
  \caption{}
  \label{fig:sub4}
\end{subfigure}\\
\begin{subfigure}{.124\textwidth}
  \centering
  \includegraphics[width=\linewidth]{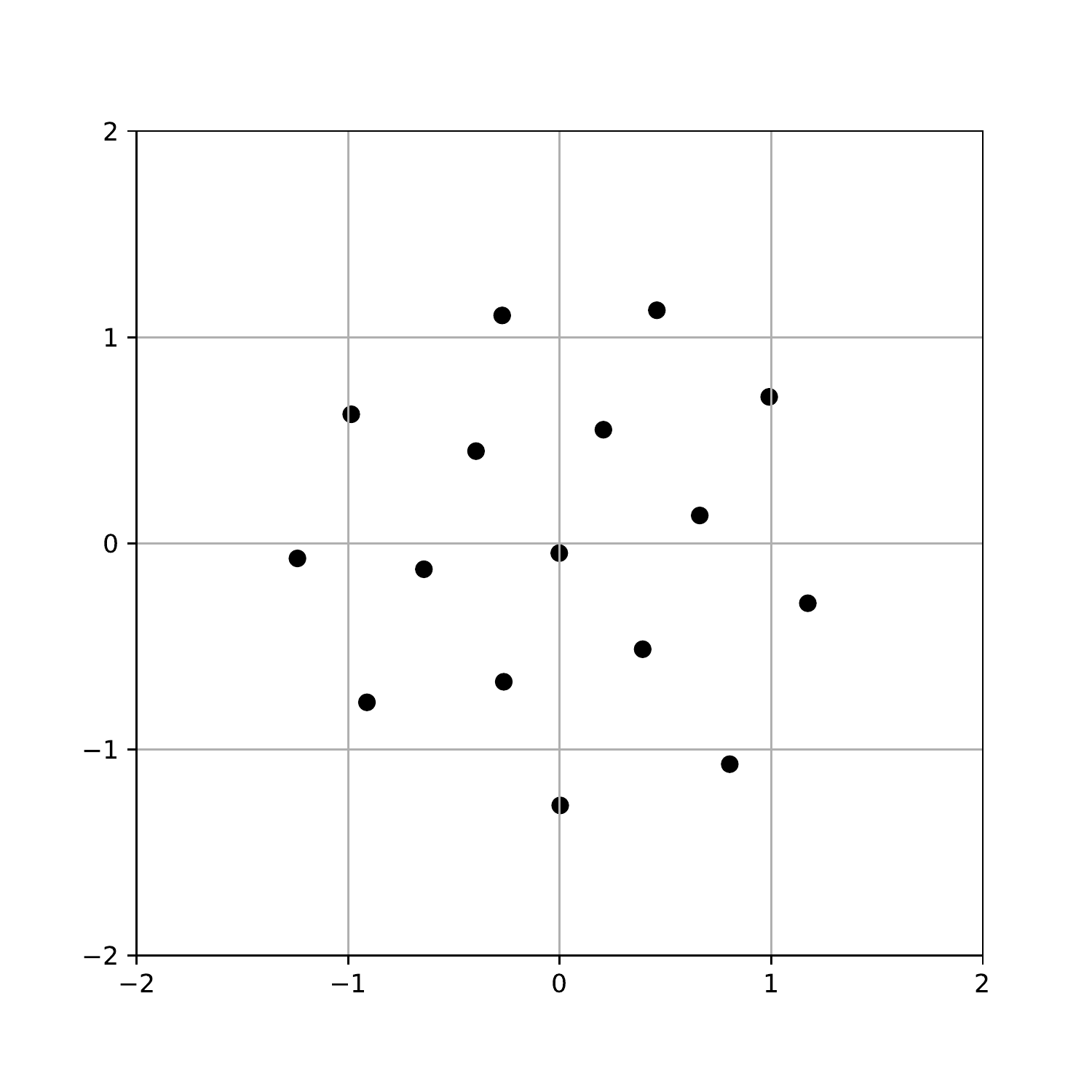}
  \caption{}
  \label{fig:sub5}
\end{subfigure}%
\begin{subfigure}{.125\textwidth}
  \centering
  \includegraphics[width=\linewidth]{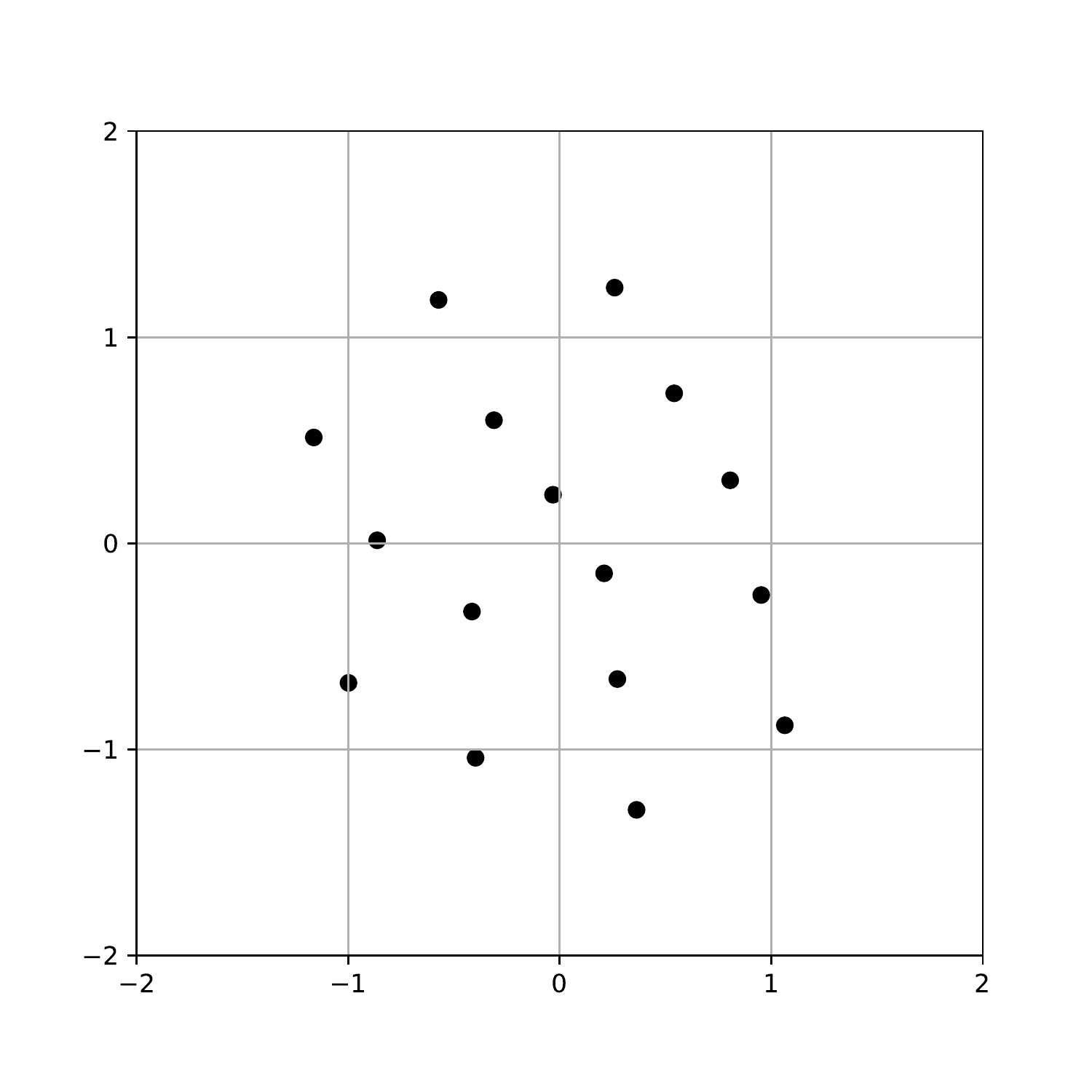}
  \caption{}
  \label{fig:sub6}
\end{subfigure}%
\begin{subfigure}{.125\textwidth}
  \centering
  \includegraphics[width=\linewidth]{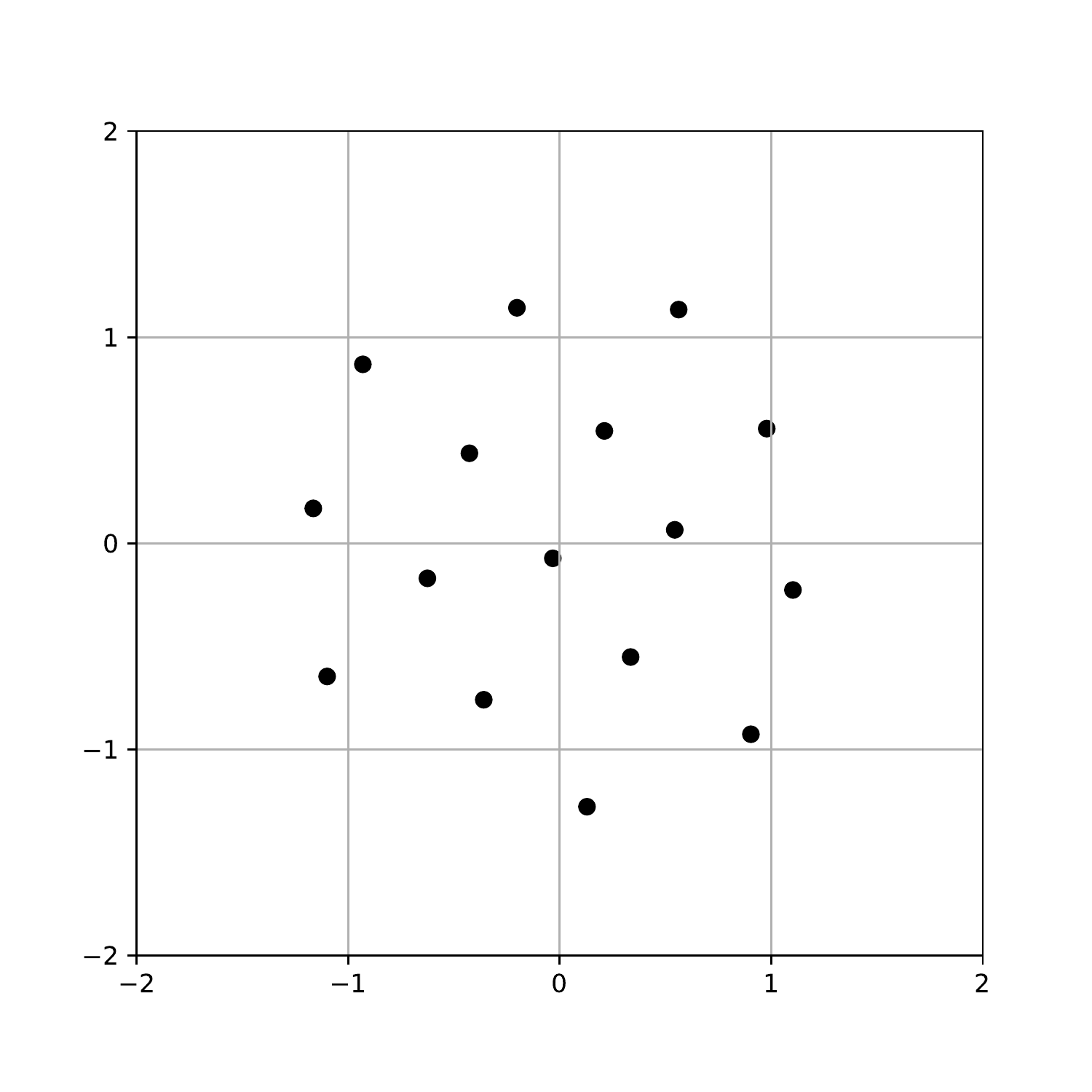}
  \caption{}
  \label{fig:sub7}
\end{subfigure}%
\begin{subfigure}{.125\textwidth}
  \centering
  \includegraphics[width=\linewidth]{figure_MI6N.pdf}
  \caption{}
  \label{fig:sub8}
\end{subfigure}
\caption{The resulting constellation points for $16$ symbols are shown by varying the number of nodes in the mutual information estimation network $T_\theta$. The constellations are for the node sizes $2$(a), $4$(b), $6$(c), $8$(d), $10$(e), $12$(f), $14$(g) and $16$(h).}
\label{fig:varyingNodes}
\end{figure}

Our encoder architecture is modelled as in \cite{dorner2018deep}, i.e., it consists of an embedding layer, dense hidden layers, converts $2n$ real values to $n$ complex values and normalizes them, see Fig.~\ref{fig:my_label}. However, unlike other end-to-end learning approaches, we estimate the mutual information from samples and train the encoder network by maximization of the mutual information estimation. This enables deep learning for channel encoding without explicit knowledge of the channel density function. Our mutual information estimation uses the Donsker-Varadhan estimator, see \eqref{DVapprox}. For that we maximize the estimator term
\begin{IEEEeqnarray*}{rCl}
        I_{\theta}(X^n;Y^n)&:=& \mathbb{E}_{p(x,y)}^{}[T_\theta(X^n,Y^n)]\\
        &&\qquad-\log \mathbb{E}^{}_{p(x)p(y)}[e^{T_\theta(X^n,Y^n)}]
\end{IEEEeqnarray*}
over $\theta$ with the Adam optimizer\cite{kingma2014adam} and a learning rate of $p=0.0005$. Note that we do not have access to the true joint distribution $p(x,y)$ and the marginal distributions $p(x)$ and $p(y)$. We therefore use samples of these distributions and approximate the expectations by the sample average. This yields the following estimator for $k$ samples
\begin{IEEEeqnarray}{rCl}
        \tilde{I}_{\theta}(X^n;Y^n)&:=& \frac{1}{k}\sum_{i=1}^{k}[T_\theta(x_{(i)}^n,y_{(i)}^n)]\IEEEnonumber\\
        &&\qquad-\log \frac{1}{k}\sum_{i=1}^{k}[e^{T_\theta(x_{(i)}^n,\bar{y}_{(i)}^n)}]\label{DVestimator},
\end{IEEEeqnarray}
where the $k$ samples of the joint distribution $p(x^n,y^n)$, for the first term in \eqref{DVestimator}, are produced via uniform generation of messages $m$ and sending them through the initialized encoder, which generates $X^n$ of $p(x^n,y^n)$. The corresponding samples of $Y^n$ are generated by our AWGN channel, see Section~\ref{Model_Section}, where the noise variance $\sigma^2$ is scaled such that we have a resulting signal-to-noise ratio per bit of $7$ $E_b/N_0$ [db]. Note also that the encoded signal $x^n$ has a unit average power normalization $\mathbb{E}(|X_i|^2)=1$, where the expectation is over the signal dimension and the batch size. The samples of the marginal distributions, for the second term in \eqref{DVestimator}, are generated by dropping either $x_{(i)}^n$ or $y_{(i)}^n$ from the joint samples $(x_{(i)}^n,y_{(i)}^n)$, and dropping the other in the next $k$ samples, as proposed in \cite{belghazi2018mine}. Therefore, a total batch size of $2k$ leads to $k$ samples of the joint and marginal distribution. In the estimator, $T_\theta$ represents a neural network, with two fully connected hidden layers and a linear output node, see Fig.~\ref{fig:MI_neuralEstimation}. Our estimator network uses $20$ nodes per hidden layer, because the mutual information value for our AWGN model stabilized at around $15$ nodes, see Fig.~\ref{fig:varyingNodes}. However, we remark that the MINE implementation of \cite{belghazi2018mine} uses $3$ fully connected layers with $400$ nodes per layer to produce the results for the $25$ Gaussians data set. A higher dimensionality therefore may require more nodes to produce stable results. As in \cite{belghazi2018mine}, we initialized the weights in $T_\theta$ with a low standard deviation ($\sigma=0.05$), to circumvent unstable behaviour in conjunction with the $\log$ term. The approximation \eqref{NGYapprox} is more stable in that regard, however, the outlook of a better estimate in high-dimensions motivated the use of the Donsker-Varadhan based estimator. 
Our training of the encoder is now implemented in two phases. After the initialization of the encoder weights, we train the mutual information estimation network for an initial round with $1000$ iterations and a batch size of $200$. In the second phase, we alternate between maximizing \eqref{DVestimator} over the encoder weights $\phi$ and the estimator weights $\theta$
\begin{equation*}
        \max_{\phi}\max_{\theta} \tilde{I}_{\theta}(X^n_{\phi}(m);Y^n).
\end{equation*}
We maximize over the encoder weights $\phi$ with batch sizes \{$100$, $100$, $1000$\} and \{$1000$, $10000$, $10000$\} iterations with a learning rate $\alpha$ of \{$0.01$, $0.001$, $0.001$\}, respectively. After every \{$100$, $1000$, $1000$\} iterations, i.e., $10$ times during every cycle, we maximize over the estimator weights $\theta$ again with a batch size of $200$.

\subsection{Decoder Training via Cross-Entropy}
To evaluate our new method, we use a standard cross-entropy based NN decoder, consisting of a conversion from complex to real values, followed by dense hidden layers, a softmax layer and an $\arg\max$ layer, see Fig.~\ref{fig:my_label}. Let $\nu^{|\mathcal{M}|}$ be the output of the last dense layer in the decoder network. The softmax function takes $\nu^{|\mathcal{M}|}$ and returns a vector of probabilities for the message set, i.e., $p^{|\mathcal{M}|}\in (0,1)^{|\mathcal{M}|}$, where the entries $p_m$ are calculated by
\begin{equation*}
    p_m=f(\nu^{|\mathcal{M}|})_m:=\frac{\exp(\nu_m)}{\sum_i \exp(\nu_i)}.
\end{equation*}
The decoder then declares the estimated messages to be $\hat{m}=\arg\max_m p_m$. For the training, we uniformly generate message indexes $m$, and feed them in our previously trained encoder NN, which generates the codewords $x^n(m)$. The codewords are then sent over the channel. The receiver gets a noisy signal $y^n$ and feeds it into the decoder. The decoder outputs the estimated probabilities $p_m$ of the received message index and feeds it into a cross-entropy function together with the true index $m$
\begin{IEEEeqnarray*}{rCl}
H(M,\hat{M})&=&-\sum_{m\in \mathcal{M}}p(m)\log p_{\text{decoder}}(m)\IEEEnonumber\\
&=&-\mathbb{E}_{p(m)}[\log  p_{\text{decoder}}(m)],
\end{IEEEeqnarray*} which is estimated by averaging over the sample size $k$, which yields the cross entropy cost function for the decoder weights~$\psi$
\begin{IEEEeqnarray*}{rCl}
    J(\psi)&=&-\frac{1}{k}\sum_{i=1}^k \log p_m,
\end{IEEEeqnarray*} where $m$ represents the index of the message of the $i$-th sample.
Finally, we use the Adam optimizer to train the decoder weights $\psi$ by minimizing the cross-entropy. We remark that we assume, that our decoding system knows the sent message index, this can be achieved by using a fixed seed on a random number generator, as proposed in \cite{goutay2018deep}.


\subsection{Results}

\begin{figure}
    \centering
    \includegraphics[scale=0.37]{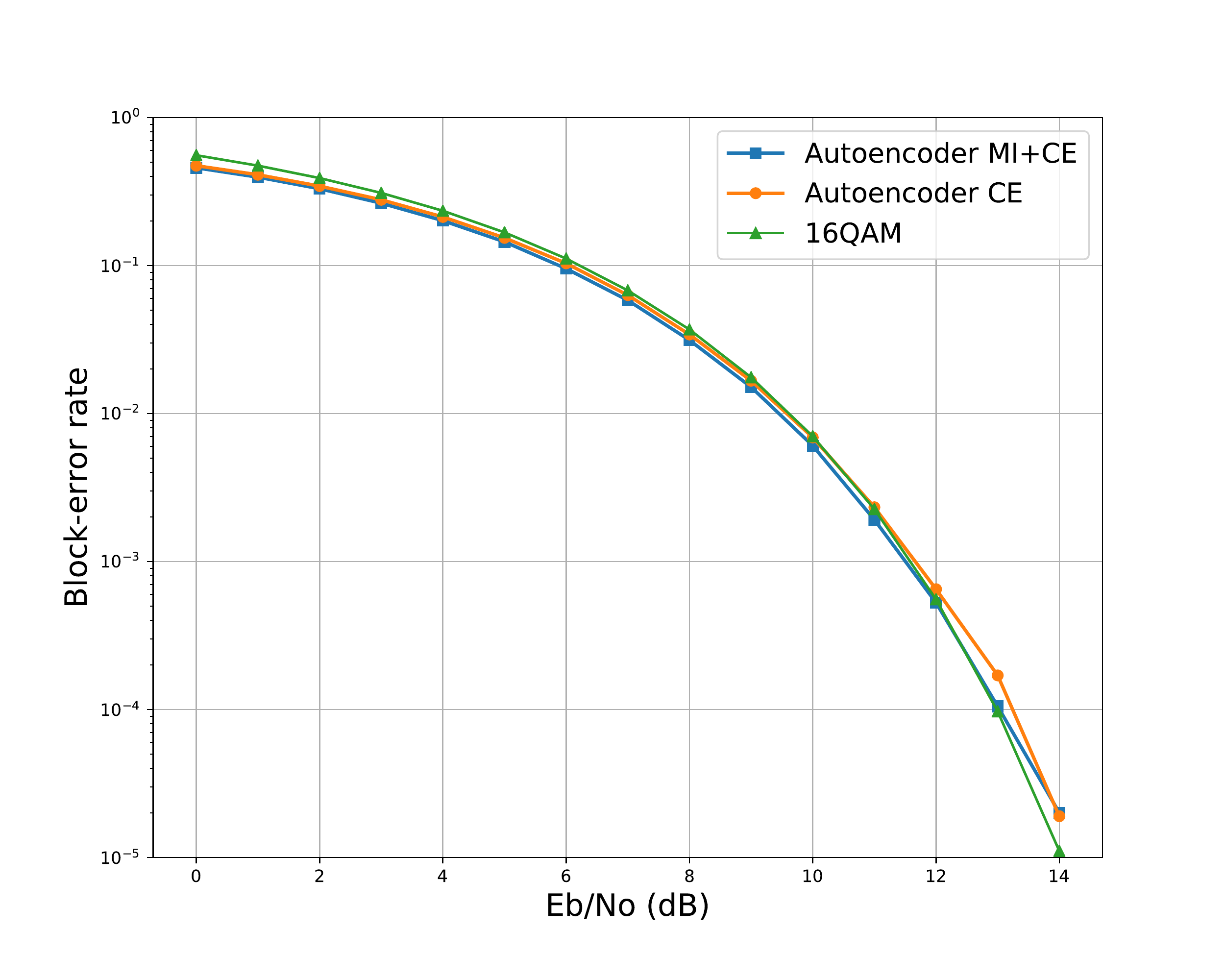}
    \caption{The resulting averaged block-error rates $P_e$ is shown for $n=1$ and the following systems: a standard $16$ QAM coding scheme; an end-to-end learning system based on cross-entropy (CE) with known channel distribution; and our proposed mutual information maximization encoding and cross-entropy decoding system (MI+CE) without known channel distribution, i.e., sample based. }
    \label{fig:BLER_simu}
\end{figure}

\begin{figure}
    \centering
    \includegraphics[scale=0.37]{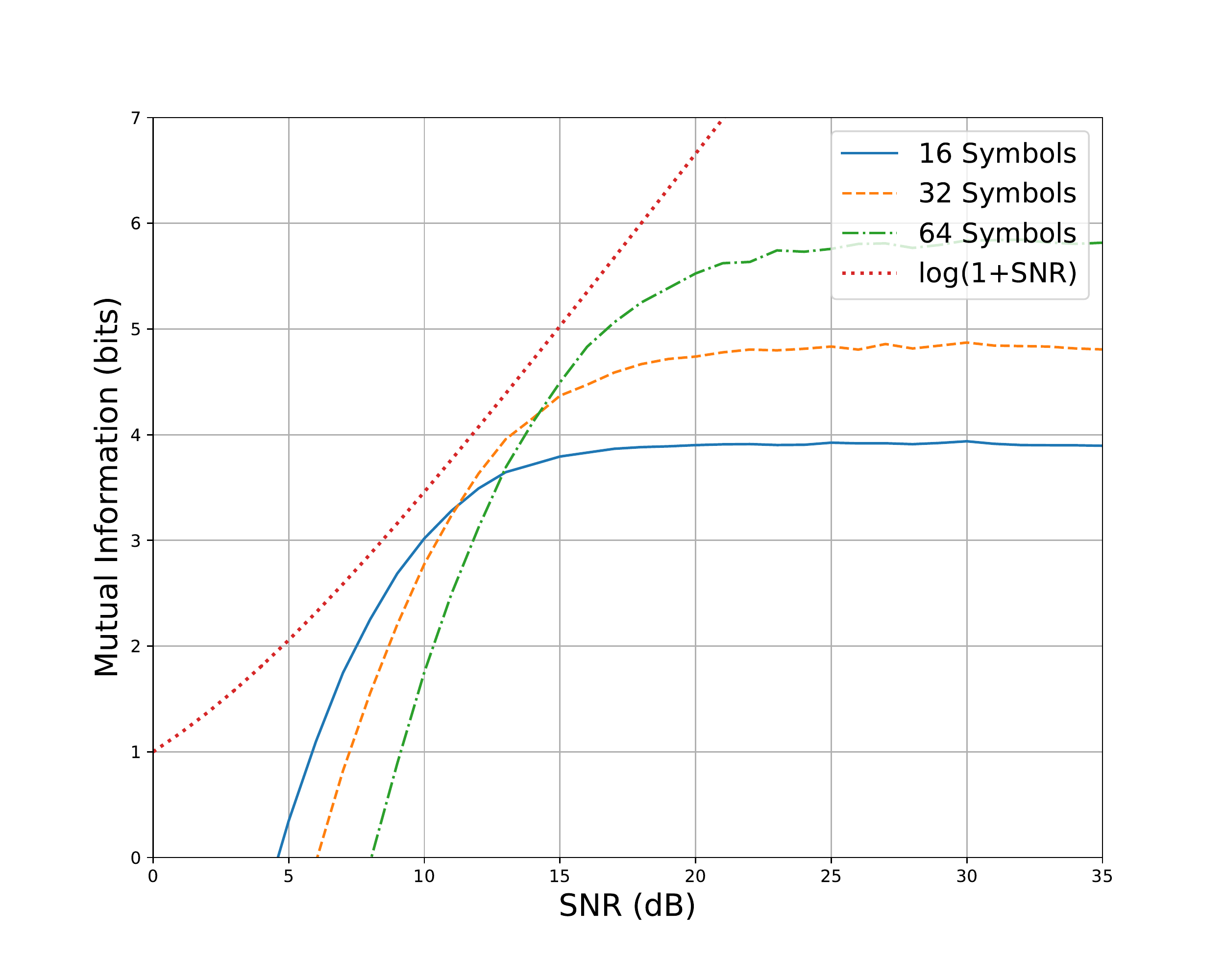}
    \caption{Mutual information estimation evaluations for trained encoders for $n=1$ and \{$16$, $32$, $64$\} symbols. The encoder and the estimation network $T_\theta$ were trained as in Section \ref{Training_Encoder} with increasing SNR values: \{$10:14$, $14:18$, $17:21$\} dB for \{$16$, $32$, $64$\} symbols, respectively.}
    \label{fig:MI_Simu}
\end{figure}

\begin{figure}
    \centering
    
\begin{subfigure}{.24\textwidth}
  \centering
  \includegraphics[width=\linewidth]{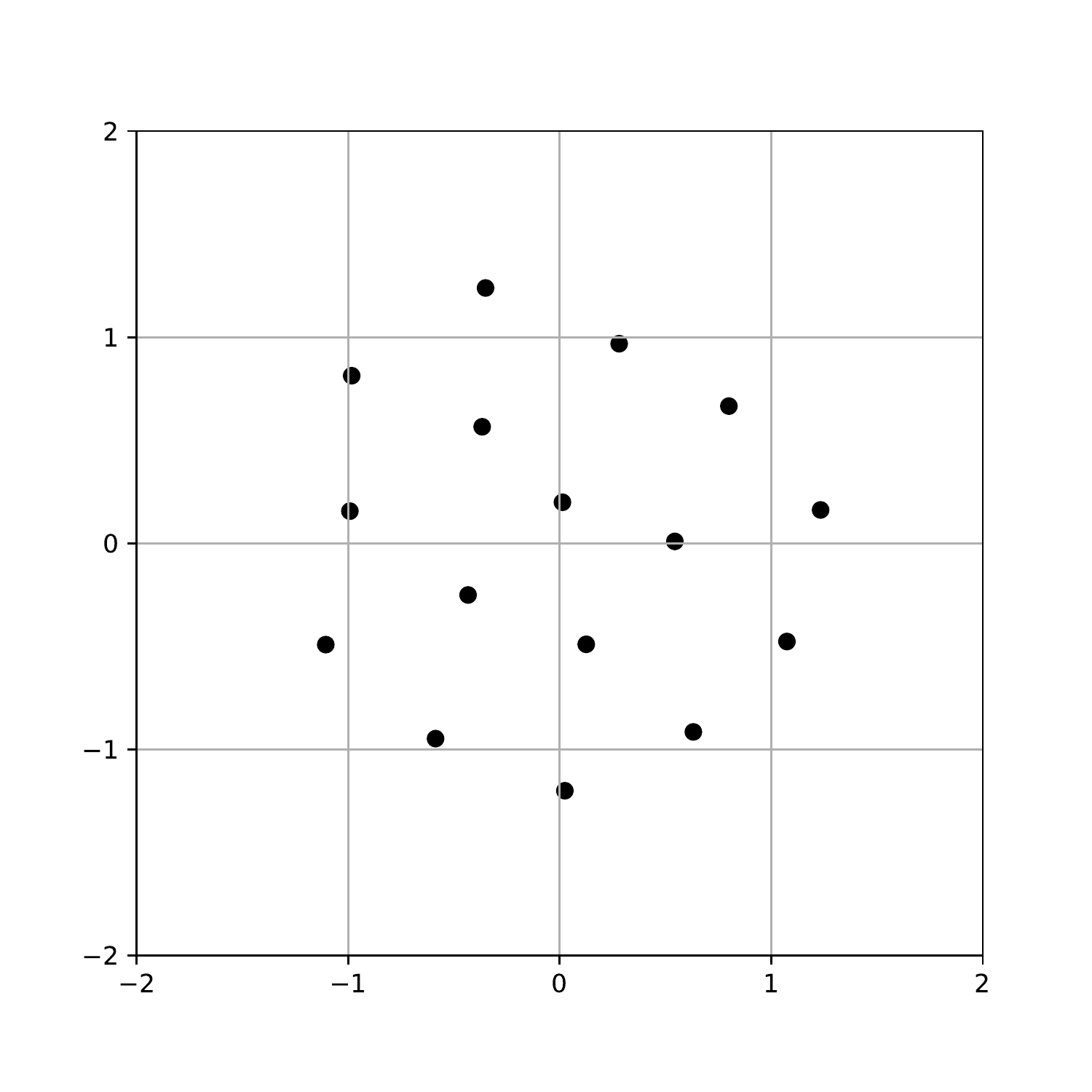}
  \caption{}
  \label{fig:sub11}
\end{subfigure}%
\begin{subfigure}{.24\textwidth}
  \centering
  \includegraphics[width=\linewidth]{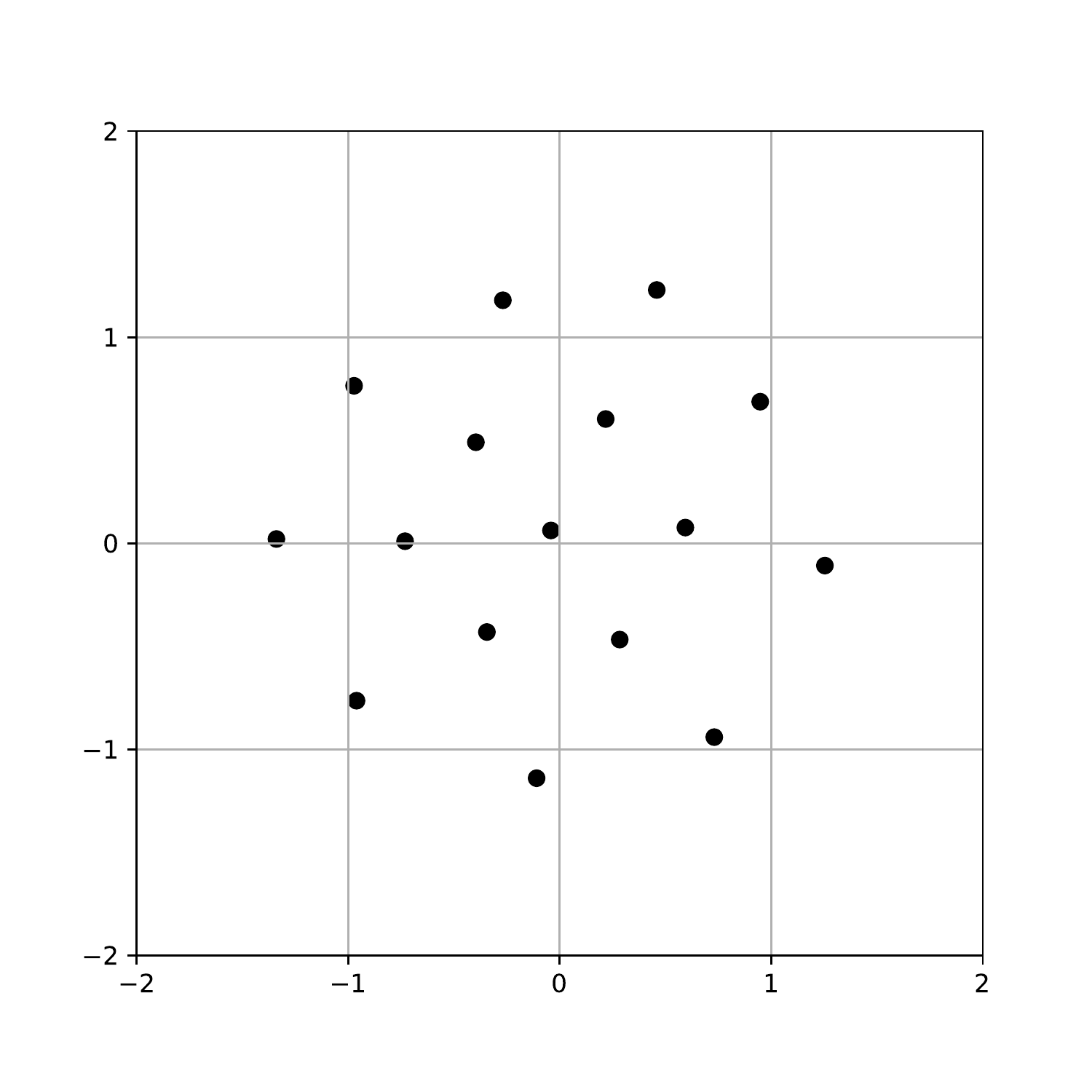}
  \caption{}
  \label{fig:sub22}
\end{subfigure}
\caption{The resulting encoding constellations are shown for $16$ symbols based on: (a) the standard cross-entropy approach; (b) the mutual information estimation based approach.}
\label{fig:Const}
\end{figure}

We have implemented our new approach using the TensorFlow \cite{tensorflow2015-whitepaper} framework. The resulting symbol error curves can be seen in Fig.~\ref{fig:BLER_simu}, which show that the performance of our proposed method is indistinguishable from the theoretical approximated 16 QAM performance and the state-of-the-art autoencoder which uses the knowledge of the channel in conjunction with a cross-entropy loss. Moreover, Fig.~\ref{fig:Const} shows the resulting constellations of the encoder for 16 symbols, for (a) the standard cross-entropy approach and (b) the mutual information maximization approach. Furthermore, we show in Fig.~\ref{fig:MI_Simu} the calculated mutual information values after training, for several signal-to-noise ratios. There, we compare these values for $16$, $32$, and $64$ symbols. We remark that the performance of the encoders is dependent on the SNR during training. If we train for example the $64$ symbols encoder at a high SNR, then the curve gets closer to $6$ bit, but also decreases in the low SNR regime. It can be seen, that the mutual information bound comes close to expected values, i.e., in the range of $m$-ary QAM, for mid to high SNR ranges.

\section{Conclusions and outlook}
We have shown that the recently developed mutual information neural estimator (MINE) can be used to train a channel encoding setup, by alternating the maximization of the estimated mutual information over the estimator weights and the encoder weights. The training works without explicit knowledge of the channel density function and rather approximates a function of the channel, i.e., the mutual information, based on the samples of the input and output of the channel. We believe that this can perform better than an end-to-end learning setup, because the encoder basically uses the expert information about which performance function (i.e. the mutual information) it needs to optimize the encoding in order to perform well. This is in contrast to the end-to-end learning approach, where the neural network system needs to learn this information on its own. Furthermore, our method can be implemented without changes at the receiver side and is therefore suitable for fast deployment. The investigation of sample size bounds for our method is on-going work. Intuitively, it requires less samples than a GAN based approach due to the fact that we do not need to exactly generate the channel density. However, future research needs to investigate the performance under low sample size scenarios and compare it to GAN and RL based approaches. Moreover, the stability of the estimator, in comparison to the $f$-divergence estimator, needs to be investigated under different channel models. 

\medskip

\small

\bibliographystyle{./IEEEtran}
\bibliography{./ref}

\end{document}